\documentclass[conference]{IEEEtran}
\IEEEoverridecommandlockouts

\usepackage{cite}
\usepackage{amsmath,amssymb,amsfonts}
\usepackage{algorithm}
\usepackage{algorithmic}
\usepackage{graphicx}
\usepackage{textcomp}
\usepackage{xcolor}
\usepackage{booktabs}
\usepackage{multirow}
\usepackage{array}
\usepackage{url}
\usepackage{graphicx}
\usepackage{booktabs}
\usepackage{tikz}
\usepackage{pgfplots}
\usepackage[normalem]{ulem}
\usepackage[hidelinks]{hyperref}

\newcommand{\clip}{\operatorname{clip}}
\newcommand{\Acc}{\operatorname{Acc}}

\newcommand{\score}{\operatorname{score}}
\newcommand{\argmax}{\operatorname*{arg\,max}}

\def\BibTeX{{\rm B\kern-.05em{\sc i\kern-.025em b}\kern-.08em
    T\kern-.1667em\lower.7ex\hbox{E}\kern-.125emX}}

\usetikzlibrary{arrows.meta,positioning,fit,shapes.geometric,calc,backgrounds,shadows}
\pgfplotsset{compat=1.16}

\begin{document}

\title{MARS: A Momentum-Aware Adaptive Rating System for Personalized Assessment}

\author{
    \IEEEauthorblockN{
        Vikrant Sahu, 
        Khethavath Sunil Naik, and 
        Gagan Raj Gupta
    }
    \IEEEauthorblockA{
        \textit{Department of Computer Science and Engineering}, 
        \textit{Indian Institute of Technology Bhilai},
        Chhattisgarh, India \\
        \{vikrantsahu, khethavathn, gagan\}@iitbhilai.ac.in
    }
}

\maketitle

\begin{abstract}
Adaptive assessment systems personalize learning by estimating learner proficiency and recommending suitable questions. However, many existing approaches rely on fixed difficulty labels, static rating updates, or black-box prediction models that provide limited control over rating stability, interpretability, and time-constrained routing. This paper proposes MARS, a Momentum-Aware Adaptive Rating System for interpretable adaptive assessment. MARS represents the question bank as a directed weighted knowledge graph and updates learner proficiency using dynamic rating sensitivity, response quality, and momentum-aware learner velocity. We also formulate a time-budgeted mark-maximization objective in which the policy selects questions within a fixed test duration under negative marking. Experiments on ASSISTments 2009 and MARS-SimExam evaluate MARS for prediction, rating stability, and adaptive routing. Results show that MARS provides smoother rating updates than Fixed Elo and achieves the highest score under time-budgeted negative-marking routing, improving the total score by 8.09 over the strongest baseline and reducing overtime violations from 50\% to 5.6\%. These results suggest that MARS is suitable for real-time, interpretable, and deployment-friendly adaptive assessment.
\end{abstract}

\begin{IEEEkeywords}
Adaptive assessment, educational data mining, Elo rating, momentum, knowledge tracing, graph-based recommendation, time-budgeted assessment, learner modeling.
\end{IEEEkeywords}

\section{Introduction}

\IEEEPARstart{A}{daptive} assessment is an important component of intelligent learning systems because it enables question difficulty, topic coverage, and feedback to be adjusted according to learner performance. In competitive examination preparation and online learning environments, learners differ in proficiency, response speed, consistency, topic-wise weakness, and recovery after mistakes. Therefore, an effective adaptive system should not only select the next question using recent accuracy or fixed difficulty labels, but also estimate evolving learner ability, support remediation, and provide interpretable recommendation logic.

Traditional computerized adaptive testing methods, such as item response theory, provide a strong psychometric foundation but often depend on calibrated item banks and offer limited flexibility when question banks evolve over time~\cite{Lord1980,vanDerLinden2000,Wainer2000}. Rating-based models provide a lightweight alternative by modeling learner--question interactions as a match between learner ability and item difficulty~\cite{Glickman1999,Glickman2008,Herbrich2007TrueSkill}. However, fixed rating updates can be unstable, particularly for new learners, and may overlook useful response-quality signals such as response time, hint usage, streak behavior, and topic structure. Knowledge tracing models, including Bayesian, deep, and graph-based variants, have shown strong performance in correctness prediction, but they often provide limited transparency for adaptive routing and remediation decisions~\cite{CorbettAnderson1994,Piech2015,Nakagawa2019}.

To address these limitations, this paper proposes \textit{MARS} (\textbf{Momentum-Aware Adaptive Rating System}), a graph-based adaptive assessment framework that models the question bank as a directed weighted knowledge graph with same-topic, prerequisite, and cross-topic relationships. MARS updates learner proficiency using dynamic rating sensitivity, response quality, and momentum-based learner velocity, enabling stable and responsive assessment. The framework supports adaptive question routing, weak-topic recovery, and interpretable remediation by selecting questions based on difficulty, learner momentum, graph structure, topic weakness, diversity, and centrality, while guiding learners through related easier concepts before revisiting unresolved topics.






The main contributions of this work are summarized as follows:
\begin{itemize}
\item \textbf{Momentum-aware adaptive rating.} We propose MARS, an interpretable online rating framework that extends Elo-style learner modeling with dynamic update sensitivity, response-quality scaling, and momentum-driven learner velocity to stabilize noisy learner-state updates.

\item \textbf{Graph-guided remediation and routing.} We design a question-graph-based routing policy that selects items using target difficulty, learner momentum, weak-topic recovery, diversity, and graph structure, enabling adaptive progression and contextual remediation.

\item \textbf{Score-aware time-budgeted assessment.} We formulate adaptive question selection as a mark-maximization problem under fixed test duration, and evaluate MARS against rating-based and knowledge-tracing baselines for prediction, stability, and deployment-oriented routing.

\end{itemize}


\section{Related Work}

Computerized adaptive testing is commonly grounded in item response theory (IRT), where learner ability and item parameters are used to estimate the probability of a correct response~\cite{Lord1980,vanDerLinden2000,Wainer2000}. Although IRT provides a strong psychometric foundation, it usually requires calibrated item banks and sufficient historical responses, which can be limiting when question banks evolve. Rating-based models provide a lightweight online alternative by treating learner--item interaction as a match between learner ability and item difficulty~\cite{Glickman1999,Herbrich2007TrueSkill}. However, fixed Elo-style updates may be unstable for new learners and may not account for response quality, topic weakness, remediation, or graph structure. Glicko-style extensions introduce uncertainty and volatility into rating estimation~\cite{Glickman2008}, but they are not designed for educational routing with response-time, hint, and graph-based signals.

Knowledge tracing models estimate learner knowledge from sequential interaction data. Bayesian Knowledge Tracing models mastery as a latent skill state updated using learning, guessing, and slipping parameters~\cite{CorbettAnderson1994,Yudelson2013}, while classical factor-based, matrix-factorization, and logistic learner models use prior success and failure counts to estimate skill acquisition~\cite{Pavlik2009,Cen2006,ThaiNghe2010,Pelanek2017}. Deep Knowledge Tracing introduced recurrent neural networks for learner-response sequences~\cite{Piech2015}, and later analyses and extensions examined deeper recurrent models, regularized sequence models, memory networks, attention, transformers, and graph neural networks to improve prediction~\cite{Xiong2016Grit,Yeung2018,Zhang2017DKVMN,Abdelrahman2019SKVMN,Pandey2019SAKT,Pandey2020RKT,Ghosh2020AKT,Choi2020SAINT,Minn2018,Vaswani2017,Nakagawa2019}. Survey work also shows that prediction gains alone do not solve interpretability and deployment constraints in adaptive assessment~\cite{Abdelrahman2023Survey,Namoun2021SLR}. These models can achieve strong next-response prediction, but their internal states are often less transparent for real-time routing, remediation, and score-aware item selection.

Graph-based knowledge tracing and educational recommendation methods use relationships among skills, concepts, or questions to improve learner modeling and recommendation~\cite{Nakagawa2019,Liu2020GIKT,Tong2020SKT,LinWang2023}. Recent work further combines graph structure, semantic item representations, and attention mechanisms for knowledge tracing~\cite{Wang2022SEEP,Liang2024GELT,XuHu2024MBQ,Yang2024TSKT,Han2024KTDeeper,Duan2024DKTMR,LuTongCheng2024AdvancedKT}. Response time and hint usage are also important indicators of response quality and can help distinguish mastery, guessing, and effortful failure~\cite{Baker2008,Khajah2016}. MARS combines these directions by maintaining explicit learner-state variables, including rating, velocity, streak, weak topics, answered items, served items, and remediation status. Unlike prediction-only knowledge tracing models, MARS uses graph structure, response quality, and momentum-aware rating updates not only for learner-state estimation but also for interpretable adaptive routing under time and scoring constraints.

\section{Problem Formulation}

\subsection{Learner-Item Interaction Setting}

Let $\mathcal{L}=\{l_1,l_2,\ldots,l_N\}$ denote a set of learners and let $\mathcal{Q}=\{q_1,q_2,\ldots,q_M\}$ denote a question bank. Each question $q \in \mathcal{Q}$ is associated with a set of metadata attributes,
\begin{equation}
    x_q = (\text{domain}_q, \text{topic}_q, R_q, m_q, t_q, \text{type}_q),
\end{equation}
where $\text{domain}_q$ is the subject or domain, $\text{topic}_q$ is the topic label, $R_q$ is the item rating or difficulty estimate, $m_q$ is the mark value, $t_q$ is the reference solve time, and $\text{type}_q$ denotes the question type.

At interaction step $p$, learner $l_i$ receives a question $q_{i,p}$ and produces a response represented as
\begin{equation}
    r_{i,p} = (q_{i,p}, y_{i,p}, \Delta t_{i,p}, h_{i,p}),
\end{equation}
where $y_{i,p}\in\{0,1\}$ is the binary correctness outcome, $\Delta t_{i,p}$ is the time spent on the item, and $h_{i,p}$ denotes hint usage or assistance level. The adaptive system observes the response and updates the learner state before recommending the next question.

\subsection{Question Knowledge Graph}

The question bank is represented as a directed weighted graph,
\begin{equation}
    G=(V,E,w),
\end{equation}
where each node $v \in V$ corresponds to a question $q \in \mathcal{Q}$, each directed edge $(u,v)\in E$ represents a pedagogical relation between two questions, and $w(u,v)\in[0,1]$ denotes the edge distance or dissimilarity. Lower edge weight indicates stronger pedagogical closeness.

The graph contains three types of edges: same-topic edges, subject-flow edges, and bridge edges. Same-topic edges connect questions from the same topic in increasing or comparable difficulty order. Subject-flow edges connect questions according to prerequisite or curriculum sequence. Bridge edges connect questions across topics or domains when conceptual dependency exists.

For two connected items $u$ and $v$, the edge weight is defined as
\begin{equation}
\begin{split}
    w(u,v) = {} & \alpha_1 d_R(u,v) + \alpha_2 d_m(u,v) \\
    & + \alpha_3 d_{\text{topic}}(u,v) + \alpha_4 d_{\text{type}}(u,v),
\end{split}
\end{equation}

subject to
\begin{equation}
    \sum_{j=1}^{4}\alpha_j = 1, \qquad \alpha_j \geq 0.
\end{equation}
Here, $d_R$ is the normalized rating distance, $d_m$ is the normalized mark distance, $d_{\text{topic}}$ is the topic distance, and $d_{\text{type}}$ is the question-type distance.

\subsection{Learner State Representation}

At each step $p$, the learner is represented by a state vector
\begin{equation}
    s_{i,p} = (R_{i,p}, v_{i,p}, n_{i,p}, \sigma^2_{i,p}, k_{i,p}, m_{i,p}, W_{i,p}, Q^{ans}_{i,p}, Q^{srv}_{i,p}, \pi^{rem}_{i,p}),
\end{equation}
where $R_{i,p}$ is the current learner rating, $v_{i,p}$ is the learner progression velocity, $n_{i,p}$ is the number of attempted questions, $\sigma^2_{i,p}$ is the recent rating variance, $k_{i,p}$ is the signed correctness streak, and $m_{i,p}$ is the behavioural momentum label. The set $W_{i,p}$ stores weak topics, $Q^{ans}_{i,p}$ stores answered items, $Q^{srv}_{i,p}$ stores served items, and $\pi^{rem}_{i,p}$ stores remediation progress.

The weak-topic set is defined as
\begin{equation}
    W_{i,p} = \{z \in \mathcal{T} \;|\; \Acc_{i,p}(z) < \tau_w \},
\end{equation}
where $\mathcal{T}$ is the set of topics, $\Acc_{i,p}(z)$ is the learner's accuracy on topic $z$, and if $\tau_w  $ is less than 40\% is the weak-topic threshold. 

\subsection{Adaptive Question Routing Objective}

The main objective is to select the next question $q^{*}_{i,p+1}$ that is suitable for the learner's current state while satisfying graph, difficulty, momentum, and remediation constraints. The recommendation problem is formulated as
\begin{equation}
    q^{*}_{i,p+1} = \argmax_{q \in \mathcal{C}_{i,p}} U(q \mid s_{i,p}, G),
\end{equation}
where $\mathcal{C}_{i,p}$ is the candidate set and $U(q \mid s_{i,p},G)$ is the adaptive utility score.

For normal progression (non-remediation), the candidate set is selected from the graph neighbourhood of the last attempted question:
\begin{equation}
    \mathcal{C}_{i,p}^{std} =
    \{q' \in V : d_G(q_{i,p},q') \leq k_h, q' \notin Q^{ans}_{i,p}, q' \notin Q^{srv}_{i,p}\},
\end{equation}
where $d_G(\cdot,\cdot)$ is graph hop distance and $k_h$ is the maximum hop limit.

For remediation, if the learner has an unresolved missed item $q_m$, the candidate set is restricted to related but easier questions:
\begin{equation}
    \mathcal{C}_{i,p}^{rem} =
    \{q' \in V : d_G(q_m,q') \leq h, q' \notin Q^{ans}_{i,p}, R_{q'} < R_{q_m}\}.
\end{equation}
This ensures that remediation questions are pedagogically connected to the missed item while being easier than the original question.



\subsection{Time-Budgeted Mark-Maximization Objective}

For the time-budgeted setting, the policy selects at most $J$ questions within a duration budget $T$. A correct response earns $m(q)$ marks, while an incorrect response receives a penalty of $\rho m(q)$. Thus, the expected score of question $q_j$ is
\begin{equation}
\begin{split}
    \hat{S}(q_j \mid s_j) = {} & \hat{P}(y_j=1 \mid s_j,q_j)m(q_j) \\
    & - (1-\hat{P}(y_j=1 \mid s_j,q_j))\rho m(q_j).
\end{split}
\end{equation}

The routing objective is
\begin{equation}
\max_{\pi}\sum_{j=1}^{J}\hat{S}(q_j \mid s_j)
\quad
\text{s.t.}
\quad
\sum_{j=1}^{J}\hat{t}(q_j \mid s_j)\leq T .
\end{equation}
In the experiments, $J=50$, $T=3600$ seconds, and $\rho=0.25$.

\subsection{Rating Update Objective}

After each response, the system updates the learner rating using an online rating function. Unlike direct Elo-style update rules, MARS first computes a raw rating update and then applies a momentum velocity update:
\begin{equation}
    \Delta R^{raw}_{i,p} = K_{i,p}\,Q_{i,p}\,(S_{i,p}-E_{i,p}),
\end{equation}
\begin{equation}
    v_{i,p} = \beta_{i,p}v_{i,p-1} + (1-\beta_{i,p})\Delta R^{raw}_{i,p},
\end{equation}
\begin{equation}
    R_{i,p+1} = R_{i,p} + \clip(v_{i,p},-v_{max},v_{max}).
\end{equation}

Here, $K_{i,p}$ is the dynamic update factor, $S_{i,p}\in\{0,1\}$ is the observed correctness, $Q_{i,p}=0.85+0.30\,P_{i,p}$ is a multiplier based on the response quality score $P_{i,p}$, $E_{i,p}$ is the expected probability of success, $v_{i,p}$ is the accumulated progression velocity, $\beta_{i,p}$ is the momentum coefficient, and $v_{max}$ prevents sudden rating jumps.

The expected probability of success is defined as
\begin{equation}
    E_{i,p} =
    \frac{1}{1 + 10^{(R^{eff}_{q_{i,p}} - R_{i,p})/400}},
\end{equation}
where $R^{eff}_{q_{i,p}}$ is the effective item rating. The effective rating may include graph centrality adjustment:
\begin{equation}
    R^{eff}_{q} = R_q + \xi c(q),
\end{equation}
where $c(q)\in[0,1]$ is the normalized centrality score of item $q$ and $\xi$ controls the maximum centrality contribution.




\begin{figure*}[ht]
\centering
\includegraphics[width=\linewidth]{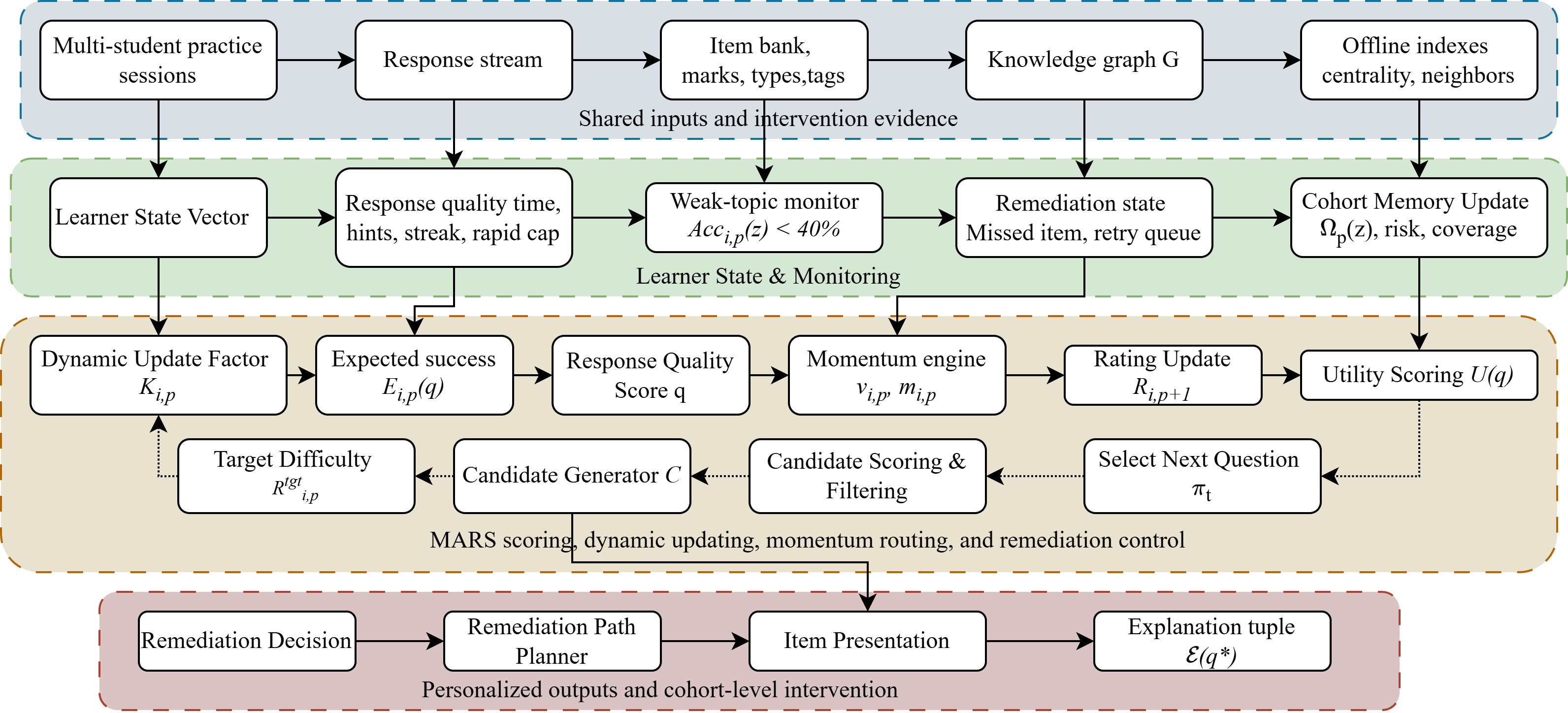}
\caption{Overview of the MARS online rating and routing pipeline.}
\label{fig:workflow}
\end{figure*}
\section{Proposed Methodology}

This section presents the proposed MARS framework. MARS extends rating-based adaptive assessment by integrating dynamic learner uncertainty, multi-signal response quality, graph-aware item calibration, momentum-inspired learner progression, rapid-response anomaly control, and contextual remediation.

\subsection{System Overview}

At each learner interaction, MARS receives the current learner state $s_{i,p}$, the attempted question $q_{i,p}$, and the observed response $r_{i,p}$. The system first computes the expected correctness probability using the learner rating and the effective item rating. It then computes a response quality score using correctness, response time, hint usage, and streak quality. This response quality is compared with the expected probability to produce a raw rating update. The raw update is passed through the momentum velocity layer, which smooths noisy individual responses and updates the learner rating. Finally, the next question is selected from either a standard graph neighbourhood or a remediation neighbourhood.

Figure~\ref{fig:workflow} illustrates the MARS adaptive assessment pipeline. Learner responses, question-bank metadata, and knowledge-graph relations are used to update learner state through expected success estimation, response-quality scoring, dynamic rating updates, and momentum-aware smoothing. The updated state then guides candidate generation and scoring using graph proximity, difficulty alignment, weak-topic recovery, diversity, and risk-aware utility. The selected question is delivered with an interpretable explanation, and the resulting response closes the feedback loop for subsequent routing and remediation.

\subsection{Behavioral Momentum-Aware Adaptive Routing}

MARS maintains a behavioural momentum state to characterize the recent learning dynamics of a learner. The behavioural state is defined as:
\begin{equation}
m_{i,p} =
\begin{cases}
\text{hot}, & k_{i,p}\geq 5 \land \hat{a}_{i,p}\geq 0.80,\\
\text{cold}, & k_{i,p}\leq -3 \land \hat{a}_{i,p}<0.40,\\
\text{steady}, & \text{otherwise},
\end{cases}
\end{equation}
where $k_{i,p}$ denotes the recent correctness streak and $\hat{a}_{i,p}$ represents the recent session accuracy of learner $i$ at interaction step $p$.

Unlike static routing strategies that apply fixed difficulty shifts, MARS dynamically estimates the target difficulty transition based on learner velocity, recent performance, and rating volatility. The target rating for the next recommended item is computed as:
\begin{equation}
R^{tgt}_{i,p}
=
R_{i,p}
+
\Delta R^{tgt}_{i,p},
\end{equation}
where the adaptive target offset is defined as:
\begin{equation}
\begin{split}
\Delta R^{tgt}_{i,p} = \operatorname{clip}\big(\alpha_v v_{i,p} + \alpha_a(\hat{a}_{i,p}-a_0) - \alpha_{\sigma}\tfrac{\sigma^2_{i,p}}{\sigma^2_{i,p}+\sigma_0^2}, \\
-\Delta_{max}, \Delta_{max}\big).
\end{split}
\end{equation}

Here, $v_{i,t}$ denotes the momentum-aware learner velocity, $\hat{a}_{i,t}$ is the recent rolling accuracy, $a_0$ is the desired target accuracy level, and $\sigma^2_{i,t}$ represents the recent rating volatility of the learner. The clipping operator constrains the target shift within the interval $[-\Delta_{max}, \Delta_{max}]$ to avoid abrupt difficulty transitions.

The behavioural momentum label is retained as an interpretable learner-state descriptor, while the actual routing transition is determined dynamically through the adaptive offset formulation. This enables MARS to provide smoother and context-aware progression across the question graph compared with static rule-based routing policies.



\subsection{Response Quality Score}

Instead of using only binary correctness, MARS computes a response quality score $P_{i,p}\in[0,1]$:
\begin{equation}
P_{i,p}=S_{i,p}\left(w_{\tau}\tau_{i,p}+w_H(1-H_{i,p})+w_s\psi_{i,p}\right)+(1-S_{i,p})\epsilon_{fail},
\end{equation}
where $S_{i,p}\in\{0,1\}$ is correctness, $\tau_{i,p}$ is response-time quality, $H_{i,p}$ is normalized hint usage, $\psi_{i,p}$ is streak quality, and $\epsilon_{fail}$ is a small failure floor. The weights satisfy
\begin{equation}
    w_{\tau}+w_H+w_s=1.
\end{equation}

The response-time quality is computed with respect to reference solve time $t_{ref}(q)$:
\begin{equation}
\tau_{i,p}=\clip\left(\frac{t_{ref}(q)-t_{spent}(q)+t_{slack}(q)}{t_{ref}(q)},0,1\right).
\end{equation}
Hint usage is normalized as
\begin{equation}
H_{i,p}=\min\left(1,\frac{n^{hint}_{i,p}}{n^{hint}_{max}}\right).
\end{equation}
The streak quality is given by
\begin{equation}
\psi_{i,p}=\frac{1}{2}\left(1+\frac{k_{i,p}}{\max(|k_{i,p}|,k_{max})}\right).
\end{equation}

\subsection{Dynamic Uncertainty-Aware Update Factor}

The dynamic update factor $K_{i,p}$ controls rating sensitivity. A high $K_{i,p}$ enables rapid calibration for new learners, while a low $K_{i,p}$ stabilizes ratings for well-calibrated learners. MARS uses a piecewise factor:
\begin{equation}
K_{i,p}= 
\begin{cases}
K_{max}, \quad n_{i,p}<n_{prov},\\
\!\begin{aligned}
&K_{min}+(K_{max}-K_{min})e^{-\lambda(n_{i,p}-n_{prov})} \\
&\quad \times \left(\frac{\sigma^2_{i,p}}{\sigma^2_{i,p}+\sigma^2_0}\right), \quad n_{i,p}\geq n_{prov}.
\end{aligned}
\end{cases}
\end{equation}

Here, $K_{max}$ is the maximum update factor, $K_{min}$ is the stable floor, $n_{prov}$ is the provisional phase length, $\lambda$ is the decay rate, and $\sigma^2_0$ is a reference volatility constant.



\subsection{Momentum-Aware Learner Progression}

The central extension of MARS is the use of a learner progression velocity. The raw update is first computed as
\begin{equation}
\Delta R^{raw}_{i,p} = K_{i,p}\,Q_{i,p}\,(S_{i,p}-E_{i,p})
\end{equation}
Then the velocity is updated as
\begin{equation}
    v_{i,p}=\beta_{i,p}v_{i,p-1}+(1-\beta_{i,p})\Delta R^{raw}_{i,p}.
\end{equation}
Finally, the learner rating is updated as
\begin{equation}
    R_{i,p+1}=R_{i,p}+\clip(v_{i,p},-v_{max},v_{max}).
\end{equation}

The momentum coefficient is adapted using rating volatility. First, the volatility ratio is computed as
\begin{equation}
\nu_{i,p}=\frac{\sigma^2_{i,p}}{\sigma^2_{i,p}+\sigma^2_0}.
\end{equation}
The coefficient is then defined as
\begin{equation}
\beta_{i,p}=\beta_{min}+(\beta_{max}-\beta_{min})(1-\nu_{i,p}).
\end{equation}
Stable learners receive a higher momentum coefficient, producing smoother progression. Volatile learners receive a lower momentum coefficient, reducing the risk of carrying noisy trends too far.

\subsection{Rapid-Response Anomaly Control}

The reference solve time is estimated as
\begin{equation}
t_{ref}(q)=\alpha_0+\alpha_1\ell_{stem}(q)+\alpha_2\ell_{opt}(q)+\alpha_3c_{idx}(q),
\end{equation}
where $\ell_{stem}(q)$ is stem length, $\ell_{opt}(q)$ is option length, and $c_{idx}(q)$ is a binary complexity indicator. A response is considered rapid when
\begin{equation}
t_{spent}(q)<\eta_t t_{ref}(q).
\end{equation}
If the response is correct and rapid, the positive update is capped:
\begin{equation}
\Delta R^{raw}_{i,p}=\min(\Delta R^{raw}_{i,p},\kappa K_{min}).
\end{equation}
This preserves correctness credit while preventing artificial rating inflation.

\subsection{Candidate Scoring Function}

For each candidate $q'\in\mathcal{C}_{i,p}$, MARS computes
\begin{equation}
\score(q')=w_hh(q')+w_gg(q')+w_vM(q'),
\end{equation}
where $h(q')$ is the heuristic fit, $g(q')$ is the graph-structural score, and $M(q')$ is the momentum-alignment score.

The heuristic component is
\begin{equation}
\begin{split}
h(q') = 1 - \frac{a_1d_R(q') + a_2d_{diff}(q')}{C_{norm}} + \beta_w\mathbb{I}[q'\in W_{i,p}] \\
+ \beta_{type}\mathbb{I}[type(q') \neq type(q_{i,p})],
\end{split}
\end{equation}
where
\begin{equation}
d_R(q')=|R^{eff}_{q'}-R^{tgt}_{i,p}|.
\end{equation}
The graph-structural score is
\begin{equation}
g(q')=\gamma_1\rho_{in}(q')+\gamma_2\rho_{out}(q')+\gamma_3c(q'),
\end{equation}
where $\rho_{in}$ and $\rho_{out}$ are normalized in-degree and out-degree, and $c(q')$ is normalized centrality.

The momentum alignment term is
\begin{equation}
M(q')=
\begin{cases}
1, & v_{i,p}>0 \land R^{eff}_{q'}\geq R_{i,p},\\
1, & v_{i,p}<0 \land R^{eff}_{q'}\leq R_{i,p},\\
0.5, & |v_{i,p}|\approx 0,\\
0, & \text{otherwise}.
\end{cases}
\end{equation}
This term aligns routing with learner trajectory. Positive velocity supports controlled advancement, while negative velocity encourages consolidation or remediation.

\subsection{Contextual Remediation and Retry}

When learner $i$ answers item $q_m$ incorrectly, the system marks it as an unresolved miss:
\begin{equation}
\pi^{rem}_{i,p}\leftarrow \operatorname{active}(q_m).
\end{equation}
During remediation, candidates are restricted to graph-related easier items. The raw update is attenuated to avoid over-penalizing the learner during recovery:
\begin{equation}
\Delta R^{rem}_{i,p}=\lambda_{rem}\Delta R^{raw}_{i,p}, \qquad 0<\lambda_{rem}<1.
\end{equation}
Remediation terminates when
\begin{equation}
n^{rem}_{i,p}\geq N_r \quad \land \quad \Acc^{rem}_{i,p}\geq A_r.
\end{equation}
After this condition is satisfied, the original missed item is queued for retry. If the retry is correct, the unresolved miss is cleared. Otherwise, remediation remains active for a later session.

\begin{algorithm}[t]
\caption{MARS Adaptive Recommendation}
\label{alg:mars_recommendation}
\begin{algorithmic}[1]
\REQUIRE $s_{i,p}$, graph $G$, last item $q_{i,p}$
\ENSURE Next question $q^*$
\STATE Compute $m_{i,p}$ and $R^{tgt}_{i,p}$
\IF{remediation active}
    \STATE Build remediation candidate set $\mathcal{C}^{rem}_{i,p}$
\ELSE
    \STATE Build graph-neighbourhood set $\mathcal{C}^{std}_{i,p}$
\ENDIF
\IF{$\mathcal{C}_{i,p}=\emptyset$}
    \STATE Use fallback candidates
\ENDIF
\FOR{$q'\in\mathcal{C}_{i,p}$}
    \STATE Compute $R^{eff}_{q'}, h(q'), g(q'), M(q')$
    \STATE Compute $\score(q')=w_hh(q')+w_gg(q')+w_vM(q')$
\ENDFOR
\STATE $q^*\leftarrow\argmax_{q'\in\mathcal{C}_{i,p}}\score(q')$
\RETURN $q^*, \mathcal{E}(q^*)$
\end{algorithmic}
\end{algorithm}

\subsection{Interpretability of Recommendation}

For every recommended question, MARS produces an explanation tuple:
\begin{equation}
\mathcal{E}(q^*)=(d_G,R^{tgt},R^{eff}_{q^*},W_{bonus},c(q^*),v_{i,p},\pi^{rem}_{i,p}).
\end{equation}
Here, $d_G$ is graph distance, $R^{tgt}$ is the target rating, $R^{eff}_{q^*}$ is effective item rating, $W_{bonus}$ indicates weak-topic prioritization, $c(q^*)$ is graph centrality, $v_{i,p}$ is progression velocity, and $\pi^{rem}_{i,p}$ indicates whether the recommendation was made during remediation.

\section{Experimental Design}

This section describes the evaluation protocol used to assess MARS.The experiments are conducted on two datasets: ASSISTments 2009 for online correctness prediction and rating stability, and \textsc{MARS-SimExam}, a persona-based synthetic assessment dataset for time-budgeted routing.

\subsection{Datasets}

ASSISTments 2009 is used as the public benchmark for online correctness prediction and rating stability~\cite{Feng2009ASSISTments,Assistments2009,Stamper2010KDDCup}. It contains learner–question interactions with correctness labels, skill tags, timestamps, hint information, and response-time information. Since public educational logs are observational, they are suitable for prediction benchmarking but limited for evaluating counterfactual adaptive routing~\cite{Choi2020EdNet}.

To evaluate adaptive question selection under realistic examination constraints, we additionally use a synthetic assessment dataset generated using a Fixed-Elo-based persona simulator. The dataset contains $277{,}500$ learner--question interactions from $300$ learners, $6{,}000$ test sessions, $1{,}091$ questions, $8$ subjects, $54$ topics, and $10$ learner personas. Each interaction includes correctness, response time, question marks, learner ratings, and topic information, enabling evaluation of adaptive routing under fixed test-duration and negative-marking constraints.

\begin{table}[t]
\centering
\caption{Dataset Summary}
\label{tab:dataset_summary}
\small
\setlength{\tabcolsep}{4pt}
\begin{tabular}{lcccc}
\toprule
Dataset & Learners & Items & Interactions & Topics \\
\midrule
ASSISTments 2009 & 4,217 & 26,688 & 346,860 & 123 \\
MARS-SimExam & 300 & 1,091 & 277,500 & 54 \\
\bottomrule
\end{tabular}
\end{table}

\subsection{Question Graph Construction}

For each dataset, a directed question graph is constructed using available skill and interaction information. Same-skill edges connect questions from the same knowledge component, while transition edges are created from consecutive learner attempts. Bridge edges connect related skill regions based on cross-topic transitions. Item difficulty is estimated from training interactions using smoothed correctness statistics or provided question Elo values where available. The graph supports centrality estimation and adaptive neighbourhood selection during routing.





\subsection{Baselines and Metrics}

MARS is compared with Majority, Fixed Elo~\cite{Glickman1999}, Bayesian Knowledge Tracing (BKT)~\cite{CorbettAnderson1994}, DKT-LSTM~\cite{Piech2015}, and Graph-based Knowledge Tracing (GKT)~\cite{Nakagawa2019}. These baselines cover majority prediction, rating-based modeling, probabilistic knowledge tracing, neural sequence modeling, and graph-based knowledge tracing.

For correctness prediction, we report AUC, accuracy, log loss, and Brier score. For rating stability, we report mean absolute update (MAU) and update standard deviation (SDU), where lower values indicate smoother learner-rating trajectories.

For time-budgeted routing, each policy selects at most $N=50$ questions within $T=3600$ seconds. We report \emph{Score}, \emph{Score/min}, raw \emph{Marks}, total \emph{Time}, and \emph{Overtime}, defined as the percentage of sessions exceeding $T$ based on the time spent on individual questions from the database. Score is computed as
\begin{equation}
\text{Score}=\sum_{q\in\mathcal{S}}\left[y_qm(q)-(1-y_q)\rho m(q)\right],
\end{equation}
where $y_q$ is correctness, $m(q)$ is item marks, and $\rho=0.25$; non-negative-marking practice tests are obtained by setting $\rho=0$.

\subsection{Implementation Details}

All models use a chronological learner-level train-test split. Learner ratings are initialized to $1500$, and item ratings are estimated only from training data to avoid leakage. MARS uses dynamic $K$, momentum-aware updates, response-quality scaling, and rapid-response capping; detailed hyperparameters and ablations are provided in the supplement.

Code is available at: \url{https://github.com/zvikrnt/MARS}.

\section{Results and Discussion}

\subsection{Correctness Prediction Results}

Table~\ref{tab:prediction_results} reports results on ASSISTments 2009. GKT achieves the strongest prediction performance, followed by DKT. This is expected because neural knowledge tracing models are optimized for sequential correctness prediction. MARS does not outperform neural models on this public observational benchmark, but it remains an interpretable online rating model with stable learner-state updates.

\begin{table}[t]
\centering
\caption{Online correctness prediction results on ASSISTments.}
\label{tab:prediction_results}
\small
\setlength{\tabcolsep}{4pt}
\begin{tabular}{lcccc}
\toprule
Model & AUC & Accuracy & LogLoss & Brier \\
\midrule
Majority & 0.5000 & 0.6883 & 0.6207 & 0.2147 \\
FixedElo & 0.7737 & 0.7430 & 0.6425 & 0.1873 \\
BKT & 0.7529 & 0.7390 & 0.5215 & 0.1758 \\
DKT & \underline{0.8648} & \underline{0.8027} & \underline{0.4017} & \underline{0.1332} \\
GKT & \textbf{0.8658} & \textbf{0.8124} & \textbf{0.3967} & \textbf{0.1298} \\
\textbf{MARS} & 0.7532 & 0.7260 & 0.6858 & 0.1987 \\
\bottomrule
\end{tabular}
\end{table}

Table~\ref{tab:synthetic_prediction_results} reports results on the synthetic dataset. In this setting, MARS achieves the highest AUC, while Fixed Elo gives slightly better accuracy, log loss, and Brier score. This shows that MARS is more competitive when the environment contains adaptive routing, learner momentum, and difficulty-aware interaction dynamics.

\begin{table}[t]
\centering
\caption{Online correctness prediction results on the synthetic dataset.}
\label{tab:synthetic_prediction_results}
\small
\setlength{\tabcolsep}{4pt}
\begin{tabular}{lcccc}
\toprule
Model & AUC & Accuracy & LogLoss & Brier \\
\midrule
Majority & 0.5000 & 0.3130 & 0.8529 & 0.3272 \\
FixedElo & \underline{0.8244} & \textbf{0.8413} & \textbf{0.4665} & \textbf{0.1301} \\
BKT & 0.7043 & 0.6024 & 0.8637 & 0.2941 \\
DKT & 0.7398 & 0.6410 & 0.6129 & 0.2132 \\
GKT & 0.6909 & 0.4475 & 0.7436 & 0.2739 \\
\textbf{MARS} & \textbf{0.8265} & \underline{0.8407} & \underline{0.4906} & \underline{0.1329} \\
\bottomrule
\end{tabular}
\end{table}

These two results indicate that MARS should not be interpreted as a replacement for neural knowledge tracing models in passive prediction benchmarks. Instead, MARS is better positioned as an online adaptive rating and routing framework for environments where interpretable learner-state updates, rating stability, and adaptive intervention are required.

For rating stability, MARS reduced the mean absolute rating update from $9.47$ to $4.95$ and the update standard deviation from $13.84$ to $7.68$ compared with Fixed Elo, corresponding to a $44.5\%$ reduction in update volatility. Fixed Elo is reported here as a reference rather than a competing baseline.

\subsection{Time-Budgeted Adaptive Routing under Negative Marking}

Table~\ref{tab:adaptive_routing} reports the time-budgeted routing results under negative marking. In this experiment, each policy selects up to $50$ questions within a $3600$ second test-duration budget. The reward is computed using positive marks for correct answers and a $25\%$ penalty for incorrect answers. Therefore, total score is lower than total raw marks when a policy selects high-risk questions that are answered incorrectly. Results are averaged over 90 evaluated session-policy pairs.

\begin{table}[t]
\centering
\caption{Time-budgeted adaptive routing.}
\label{tab:adaptive_routing}
\small
\setlength{\tabcolsep}{6pt}
\begin{tabular}{lccccc}
\toprule
Policy & Score & Score/min & Marks & Overtime & Time (s) \\
\midrule
BKT  & 75.68          & 1.36          & 93.53          & 48.9\%          & 3282.91 \\
DKT  & \underline{81.19}          & \underline{1.48}          & \textbf{98.69} & 50.0\%          & 3282.84 \\
GKT  & 66.01          & 1.19          & 86.56          & \underline{46.7}\%          & \underline{3280.63} \\
\textbf{MARS} & \textbf{89.28} & \textbf{1.70} & \underline{98.27}          & \textbf{5.6\%}  & \textbf{3132.54} \\
\bottomrule
\end{tabular}
\end{table}
MARS achieves the highest total score and score efficiency among the compared BKT, DKT, and GKT routing variants. Compared with DKT, MARS improves total score by $8.09$ points and reduces overtime violations by $44.4$ percentage points. Compared with BKT and GKT, MARS improves total score by $13.60$ and $23.27$ points, respectively. These gains show that MARS is useful not only for stable online rating, but also for practical adaptive assessment settings where the system must balance expected marks, risk, and test-duration constraints.

When negative marking is in play, the gap between total marks and total score is what actually decides performance. Although DKT obtains slightly higher raw marks, its advantage does not translate into higher final score because it incurs substantially more overtime violations. Raw correctness, on its own, tells an incomplete story; any adaptive policy worth deploying in a competitive exam setting has to weigh risk and time discipline just as heavily.

\subsection{Adaptability and Deployment Suitability}

For application-oriented deployment, MARS is designed to support low-latency online updates, small per-learner memory, and interpretable routing decisions. Table~\ref{tab:adaptability_comparison} summarizes the deployment characteristics of the compared models.

\begin{table}[t]
\centering
\caption{Adaptability Comparison}
\label{tab:adaptability_comparison}
\footnotesize 
\setlength{\tabcolsep}{1.8pt} 
\begin{tabular}{l@{\hspace{2pt}}c@{\hspace{2pt}}ccc} 
\toprule
Model & Complexity & Memory & Online & Suitable Use \\
\midrule
Fixed Elo & $O(1)$ & Low & $O(1)$ & Web rating engine \\
BKT & $O(S)$ & Low & $O(1)$ & Skill mastery tracking \\
DKT & $O(Th^2)$ & High & $O(h^2)$ & GPU/server prediction \\
GKT & $O(|E|h)$ & High & $O(|E|h)$ & Graph KT server \\
\textbf{MARS} & $O(d^{k_h})$ & Low & $O(1)$ & Adaptive learning platform \\
\bottomrule
\end{tabular}
\end{table}

MARS updates learner state using only the current rating, velocity, recent accuracy, update variance, and item metadata. Therefore, it avoids full-sequence recomputation during inference and is suitable for web-based practice platforms, LMS-integrated quizzes, and mobile adaptive assessment systems.





\subsection{Overall Discussion}

The results show that MARS is not intended to replace high-capacity knowledge tracing models for passive next-response prediction: on ASSISTments 2009, DKT and GKT achieve stronger predictive performance. Instead, MARS is better positioned as an online adaptive rating and routing framework. In the time-budgeted negative-marking setting, MARS achieves the highest total score and score efficiency, while DKT obtains slightly higher raw marks but much higher overtime violations. This indicates that fixed-duration adaptive assessment requires more than maximizing correctness or raw marks; the policy must jointly consider expected correctness, mark value, penalty risk, and available time. Overall, MARS is best suited for real-time, interpretable adaptive assessment where learner states must be updated online and routing decisions must remain auditable.

\section{Limitations and Responsible Use}

MARS depends on the quality of the question graph, skill labels, and item-difficulty estimates; weak graph edges may degrade routing quality. ASSISTments 2009 is observational and cannot fully evaluate counterfactual routing, while the synthetic persona-based dataset cannot fully represent real learner behaviour. Since the synthetic dataset is generated using a Fixed-Elo-based simulator, the score-efficiency gains should be validated on real adaptive deployment logs. In preliminary routing experiments, a reward-only policy without negative marking tended to over-select high-mark, high-difficulty items, revealing an objective-misalignment risk; therefore, the final routing objective includes negative marking and time penalties to discourage risk-seeking item selection.

The present time-budgeted comparison focuses on BKT, DKT, GKT, and MARS under a common routing objective. Future work will compare MARS with uncertainty-aware Glicko-style knowledge tracing~\cite{Glickman2008}, Transformer-based knowledge tracing~\cite{Choi2020SAINT}, and recent graph-aware knowledge-tracing models~\cite{Liang2024GELT,Yang2024TSKT,Han2024KTDeeper}. Response-time features may also be affected by device, network, reading-speed, or accessibility differences. MARS should therefore be used as a decision-support tool with transparent recommendations, privacy-preserving data handling, bias auditing, and human oversight before high-stakes deployment~\cite{Barocas2019,BakerHawn2022}.

\section{Conclusion}
This paper introduces MARS (Momentum-Aware Adaptive Rating System), a personalized adaptive assessment framework that integrates learner-rating updates, response-aware proficiency estimation, graph-based question selection, momentum-driven learner-state stabilization, and interpretable recommendations. MARS is designed to provide transparent, adaptive, and real-time question selection for personalized learning environments.



In a time-budgeted negative-marking setting, MARS achieves higher total score and score efficiency than the routing baselines while substantially reducing overtime violations. These results position MARS as an interpretable, deployment-friendly adaptive routing framework rather than a replacement for neural knowledge tracing models in passive prediction benchmarks. Future work will validate MARS on real adaptive deployment logs and incorporate explicit learner uncertainty.

\clearpage

\appendix

\begin{table*}[t]
\centering
\caption{Main MARS hyperparameters used in the time-budgeted routing experiment.}
\label{tab:hyperparameters}
\scriptsize
\setlength{\tabcolsep}{3pt}
\resizebox{\textwidth}{!}{%
\begin{tabular}{llll}
\toprule
Symbol & Value & Meaning & Practical effect \\
\midrule
$w_{\text{score}}$ & 7.83 & Expected score weight & Prioritizes questions with high expected score under negative marking. \\
$w_{\text{score/time}}$ & 0.55 & Score-per-time weight & Rewards questions that provide more expected score per second. \\
$w_{\text{diff}}$ & 1.98 & Difficulty-fit weight & Encourages questions near the learner's adaptive target difficulty. \\
$w_{\text{topic}}$ & 0.80 & Weak-topic bonus & Increases priority for topics where the learner shows weakness. \\
$w_{\text{div}}$ & 1.14 & Diversity weight & Discourages repeatedly selecting questions from the same topic region. \\
$w_{\text{time}}$ & 0.31 & Time penalty weight & Penalizes questions with high predicted response time. \\
$w_{\text{overtime}}$ & 20.89 & Overtime penalty weight & Strongly discourages selections likely to exceed the test-duration budget. \\
$r_{\text{risk}}$ & 2.78 & Risk-aversion weight & Penalizes high-risk questions under negative marking. \\
$\alpha_v$ & 2.78 & Velocity routing weight & Uses learner momentum to adjust target difficulty. \\
$\alpha_a$ & 54.32 & Recent-accuracy routing weight & Shifts target difficulty according to recent learner accuracy. \\
$\alpha_{\sigma}$ & 26.31 & Volatility routing weight & Reduces difficulty jumps for unstable learner trajectories. \\
$\Delta_{\max}$ & 195.62 & Maximum target shift & Caps the adaptive difficulty shift. \\
$\beta_{\min}$ & 0.07 & Minimum momentum coefficient & Allows faster response to volatile learner behaviour. \\
$\beta_{\max}$ & 0.12 & Maximum momentum coefficient & Smooths learner-rating updates during stable behaviour. \\
$v_{\max}$ & 17.25 & Maximum velocity step & Caps the magnitude of one momentum-based rating update. \\
$K_{\max}$ & 45.0 & Maximum rating update factor & Enables faster calibration during early learner interactions. \\
$K_{\min}$ & 12.0 & Minimum rating update factor & Provides a stable lower bound for rating updates. \\
$n_{\text{prov}}$ & 30 & Provisional phase length & Number of early attempts with stronger calibration. \\
$\sigma_0^2$ & 100.0 & Reference volatility constant & Normalizes learner rating variance in update and routing rules. \\
$a_0$ & 0.70 & Target accuracy & Desired challenge level for adaptive routing. \\
$\tau_w$ & 0.40 & Weak-topic threshold & Topics below this accuracy are treated as weak topics. \\
$W_{\text{recent}}$ & 5 & Recent-topic window & Tracks recent topic usage for diversity and weak-topic control. \\
\bottomrule
\end{tabular}
}
\end{table*}

\section*{Synthetic Dataset Generation}

The MARS-SimExam, a persona-based synthetic assessment dataset was generated from a master question bank using a Fixed-Elo interaction simulator. The simulator creates $300$ learners across $10$ personas, with $30$ learners per persona. Each persona defines a different starting rating range, latent skill level, growth rate, response speed, carelessness level, challenge preference, volatility, and accuracy bias. For each learner, daily practice sessions are generated over multiple active days, and questions are selected near a target rating derived from the learner's current rating, persona challenge preference, topic continuity, and occasional remediation or retry events. Correctness is sampled from an Elo-style probability based on latent learner skill and question Elo, with additional persona-specific bias, fatigue, carelessness, and random noise. Learner ratings are then updated using a fixed Elo update with $K=32$:
\begin{equation}
\Delta R = \operatorname{round}\left(K\left(y - \frac{1}{1+10^{(R_q-R_i)/400}}\right)\right).
\end{equation}
The simulator also records response time, graph mode, edge type, remediation status, response quality, streak state, weak topic, rating before and after the response, and predicted correctness probability. The final dataset contains $277{,}500$ learner-question interactions, together with student-level and persona-level summary files.

\section*{Computational Complexity Analysis}
\label{app:complexity}

To demonstrate the suitability of the proposed MARS framework for real-time deployment in large-scale adaptive assessment environments, we analyze the online time complexity of a single learner-item interaction step. The overall time complexity is determined by three primary operations: rating state updates, candidate generation, and candidate scoring.

\subsection{Learner State and Rating Update}
Upon receiving a learner's response, MARS computes the effective item rating, expected correctness, response quality score, dynamic update factor, and momentum velocity. As defined in Equations (11) through (31), these operations consist strictly of scalar arithmetic, conditional checks, and basic mathematical functions (e.g., exponentials). Consequently, the time complexity for updating the learner state is constant:
\begin{equation}
    \mathcal{O}_{update} = \mathcal{O}(1)
\end{equation}

\subsection{Candidate Generation}
For standard progression, MARS generates a candidate set $\mathcal{C}_{i,t}^{std}$ by performing a breadth-first search (BFS) on the directed knowledge graph $G = (V, E, w)$ starting from the current item $q_{i,t}$, bounded by a maximum hop limit $k_h$. Assuming an average node out-degree of $d$, the maximum number of items explored is bounded by $d^{k_h}$. The contextual remediation candidate generation follows a similar graph traversal logic. Thus, the time complexity for candidate generation is:
\begin{equation}
    \mathcal{O}_{gen} = \mathcal{O}(d^{k_h})
\end{equation}

\subsection{Candidate Scoring and Selection}
For each candidate $q^{\prime} \in \mathcal{C}_{i,t}$, MARS computes a composite utility score comprising heuristic fit, graph-structural importance, and momentum alignment (Equation 35). Because structural metrics such as in-degree, out-degree, and node centrality are precomputed offline, scoring each candidate requires $\mathcal{O}(1)$ operations. Scoring the entire generated candidate set scales linearly with the size of the set:
\begin{equation}
    \mathcal{O}_{score} = \mathcal{O}(|\mathcal{C}_{i,t}|) \subseteq \mathcal{O}(d^{k_h})
\end{equation}

\subsection{Overall Online Complexity}
The total time complexity per interaction step is the sum of the individual operations:
\begin{equation}
    \mathcal{O}_{total} = \mathcal{O}(1) + \mathcal{O}(d^{k_h}) + \mathcal{O}(d^{k_h}) = \mathcal{O}(d^{k_h})
\end{equation}
Because the maximum hop limit $k_h$ is a small constant (e.g., $k_h \in \{1, 2, 3\}$), the online inference time scales independently of the total item bank size $|V|$. This sub-linear scaling confirms that MARS satisfies the stringent low-latency requirements of web-based and real-time adaptive learning platforms.
In Table~\ref{tab:adaptability_comparison}, \emph{Online} is the incremental cost of updating a cached learner state after one response. BKT is $O(1)$ because a response updates only the mastery of the item's own skill through a closed-form Bayes rule, even though its full state spans all $S$ skills. DKT and GKT advance an $h$-dimensional hidden state; MARS updates a few scalars.

\section*{Extra Results}
\subsection{Rating Stability}
On the ASSISTments dataset, MARS yielded substantially more stable learner-rating trajectories than Fixed Elo. The mean absolute rating update decreased from 9.47 to 4.95, and the update standard deviation decreased from 13.84 to 7.68, representing a 44.5\% reduction in rating volatility. These results suggest that MARS provides smoother learner-state estimation while preserving the interpretability and online update characteristics of rating-based adaptive assessment methods.





\subsection{Ablation Results}

Table~\ref{tab:ablation_results} presents the component-level ablation results. MARS-Final achieves a balanced trade-off between predictive performance and rating stability. Removing momentum slightly reduces AUC and accuracy while increasing update variability. Replacing dynamic $K$ with fixed $K$ produces the strongest AUC among MARS variants, but it also substantially increases update magnitude and volatility, making the learner trajectory less stable.

\begin{table}[htbp]
\centering
\caption{Ablation study of MARS components ($N = 104,858$ for all variants). $MAU$ and $SDU$ denote Mean Absolute Update and Update Standard Deviation, respectively.}
\label{tab:ablation_results}
\scriptsize
\setlength{\tabcolsep}{3.5pt} 
\begin{tabular}{lcccccc}
\toprule
Model Variant & AUC & Acc. & LogLoss & Brier & MAU & SDU \\
\midrule
MARS-Final & 0.7532 & 0.7260 & 0.6858 & 0.1987 & 4.9461 & 7.6794 \\
w/o Momentum & 0.7504 & 0.7232 & 0.6892 & 0.1999 & 4.7384 & 7.7633 \\
w/ Fixed K & \textbf{0.7698} & \textbf{0.7377} & \textbf{0.6654} & \textbf{0.1916} & 9.3175 & 13.6527 \\
w/o QualityFactor & 0.7529 & 0.7247 & 0.6875 & 0.1993 & \textbf{4.9353} & \textbf{7.6716} \\
w/o RapidCap & 0.7566 & 0.7314 & 0.6728 & 0.1949 & 5.1348 & 7.8485 \\
w/ StreakMultiplier & 0.7244 & 0.7325 & 1.0238 & 0.2016 & 5.3362 & 8.8159 \\
\bottomrule
\end{tabular}
\end{table}

The ablation results show three important findings. First, fixed $K$ improves prediction but sacrifices stability, confirming that dynamic update sensitivity is useful when stable learner trajectories are desired. Second, the momentum term provides a modest but consistent improvement over the non-momentum variant. Third, the streak multiplier substantially reduces AUC and increases log loss, indicating that streak should not directly multiply the rating update. Instead, streak is better used as a behavioural routing signal for identifying hot, steady, or cold learner states.

\section*{Metric Interpretation}

AUC measures how well a model ranks correct responses above incorrect responses and therefore reflects discrimination independent of a fixed threshold. Accuracy measures the proportion of correct binary predictions. Log loss penalizes confident but incorrect predictions more heavily and is useful for evaluating the quality of predicted probabilities. Brier score measures the average squared difference between predicted probabilities and observed outcomes, with lower values indicating better calibration. For rating stability, MAU represents the average size of rating updates after each interaction, whereas SDU measures the variability of those updates; lower values indicate smoother learner-state trajectories and lower sensitivity to noisy responses. In the time-budgeted routing setting, Score represents the final reward after penalties, Marks represents the raw marks obtained before penalties, Score/min measures time efficiency, and Overtime indicates how often a policy exceeds the allotted test duration from the dataset. It is accounted identically for every policy: item durations come from one median response-time model fitted on the training split, and an item is admitted only while elapsed time is below $T$, so a session can overshoot only on its final item. Overtime differences therefore reflect item selection, not accounting.

\section*{Hyperparameter Selection}
Sessions were split by test into disjoint train, validation and final sets in a $70/15/15$ ratio. The routing weights in Table~\ref{tab:hyperparameters} were tuned on the validation split only, using $50$ trials of Tree-structured Parzen Estimator search that maximise $\text{Score}+0.5\,\text{Score/min}-5\,\text{Overtime}-0.01\,t_{\text{over}}$, where $t_{\text{over}}$ is mean overtime in seconds. The selected weights were then frozen and evaluated once on the final split, and the same weights are used for every routing policy.

\section*{Code Availability}

The source code, experiment scripts, configuration files, and
reproduction instructions are available at
\url{https://github.com/zvikrnt/MARS}. The repository includes scripts for dataset preprocessing, online prediction, rating-stability evaluation, synthetic data generation, time-budgeted routing, and baseline comparison.

\end{document}